\documentclass[twocolumn, prb, showpacs,superscriptaddress,floatfix]{revtex4}
\usepackage{amssymb}
\usepackage{mathrsfs}
\usepackage{graphicx}
\usepackage{float}
\begin{document}

\title{Topological phases of generalized Su-Schrieffer-Heeger models}
\author{Linhu Li}
\affiliation{Beijing National
Laboratory for Condensed Matter Physics, Institute of Physics,
Chinese Academy of Sciences, Beijing 100190, China}
\author{Zhihao Xu}
\affiliation{Beijing National
Laboratory for Condensed Matter Physics, Institute of Physics,
Chinese Academy of Sciences, Beijing 100190, China}
\author{Shu Chen}
\email{schen@aphy.iphy.ac.cn}
\affiliation{Beijing National
Laboratory for Condensed Matter Physics, Institute of Physics,
Chinese Academy of Sciences, Beijing 100190, China}
\begin{abstract}
We study the extended Su-Schrieffer-Heeger model with both the nearest-neighbor and next-nearest-neighbor hopping strengths
being cyclically modulated and find the family of the model system exhibiting topologically nontrivial phases, which can
be characterized by a nonzero Chern number defined in a two-dimensional space spanned by the momentum and modulation parameter.
It is interesting that the model has a similar phase diagram as the well-known Haldane's model. We propose to use
photonic crystal systems as the idea systems to fabricate our model systems and probe their
topological properties. Some other models with modulated on-site potentials are also found to exhibit similar phase diagrams and their connection to the Haldane's model is revealed.
\end{abstract}
\pacs{05.30.Fk, 03.65.Vf, 73.21.Cd }
\maketitle
\date{today}

\section{Introduction}
The Su-Schrieffer-Heeger (SSH) model is a standard tight-binding model with spontaneous dimerization proposed by Su, Schrieffer and
Heeger to describe the one-dimensional (1D) polyacetylene \cite{SSH}. Despite of its simplicity, it has attracted
extensive studies in the past decades as it exhibits rich physical phenomena, such as topological soliton excitation, fractional charge and nontrivial edge states \cite{soliton,JR,RMP,Ruostekoski}. Recent studies of topological insulators \cite{TI} lead to reexamination of the SSH model, which belongs to the BDI symmetry class as the simplest example of 1D topological insulators \cite{Ludwig,symmetryclass}. It was found that various systems, including the graphene ribbon \cite{Delplace}, the p-orbit optical ladder system \cite{Li} and the off-diagonal bichromatic optical lattice \cite{Sun}, can be mapped into the SSH model. The
extended SSH models with periodic modulation of chemical potential were also used to study the topological charge pumping \cite{Thouless,Rice,Fu,WangL}. On the other hand, the 1D superlattice models have recently attracted great attentions as the families of superlattice systems were found to be topologically nontrivial \cite{Shu,Kraus} with the periodical modulation parameter providing an additional dimension.
These studies inspired the theoretical connection between the 1D superlattice system and the two-dimensional (2D) quantum Hall
effect of Hofstadter lattices \cite{Shu,Kraus,Zhu,Hofstadter}.

The Haldane's model \cite{Haldane} is a prototype model which may realize anomalous quantum Hall effect in a 2D honeycomb lattice by introducing a complex next-nearest-neighbor (NNN) hopping term but without any net magnetic flux through a unit cell of the system. By varying the hopping phase and the sublattice potential difference, the system can undergo a topological phase transition from a normal insulator
to a Chern insulator \cite{Haldane,Vanderbilt,Hao}. However, it is hard to realize the Haldane's model experimentally in ordinary condensed
matter systems because of the specifically staggered magnetic flux introduced in the model \cite{Shao}. The recent progress in simulation of topological phases in reduced dimensions raises the question whether a similar phase diagram of Haldane's model is realizable in 1D systems.

In this work, we shall explore the realization of Haldane's phase diagram in reduced 1D lattices by considering the extended SSH model with both the nearest-neighbor (NN) and NNN hopping amplitudes cyclically modulated by an additional parameter. As the cyclical modulation parameter is continuously changed, we demonstrate that nontrivial edge states emerge due to the introduction of specific NNN hopping terms. The additional modulation parameter and the momentum permit us to define a topological invariant in the extended 2D parameter space to distinguish different topological phases, which gives rise to a phase diagram similar to that of the 2D Haldane's model. The topologically nontrivial edge states in our model may be observed in photonic crystal systems. Furthermore, we find that an extended SSH model with the sublattice potential modulated also exhibits nontrivial topological phases.

\section{Models and results}
\subsection{Model Hamiltonian}
We consider an 1D lattice with
one unit cell made of A and B sites described by the Hamiltonian
\begin{equation}
H=H_{SSH} + H_{NNN}, \label{H1}
\end{equation}
with
\begin{eqnarray}
H_{SSH}  &=& \sum\limits_{n} \left[ t_1 c_{A,n}^{+}c_{B,n} + t_2 c_{A,n+1}^{+}c_{B,n} \right] +H.c. ,\\
H_{NNN} &=& \sum\limits_{n} \left[ t_{A}c_{A,n}^{+}c_{A,n+1}+t_{B}c_{B,n}^{+}c_{B,n+1} \right] + H.c. ,
\end{eqnarray}
where $t_1= t(1+ \delta \cos \theta) $ and $t_2= t(1 - \delta \cos \theta)$ with
$\delta$ representing the dimerization strength and $\theta$ being a cyclical parameter which can vary from $0$ to $2\pi$ continuously,
$c_{n,A}$ (or $c_{n,B}$) are the annihilation operators localized on
site A (or B) of the $n$-th cell,
and $t_{A}$ and $t_{B}$ are the NNN hopping amplitudes along sublattice A and B
respectively. Here $t=1$ is set as the unit of energy.
When $t_{A}=t_{B}=0$, the Hamiltonian reduces to the SSH model \cite{SSH}.

Under the periodic boundary condition, we can make a Fourier
transformation: $c_{\alpha, n}=1/\sqrt{N}\sum\limits_{k}e^{ikn}c_{\alpha, k}$,
where $N$ is the number of the unit cells and $\alpha$ takes $A$ or $B$. Then the Hamiltonian can be written
in the form of
\begin{equation}
H = \psi^{\dagger}_k h(k) \psi_k ,
\end{equation}
where $\psi^{\dagger}_k = (c^{\dagger}_{A, k},c^{\dagger}_{B, k})$ and
\begin{eqnarray}
h(k)=
\left(
\begin{array}{cc}
2t_{A}\cos k & t_1 + t_2 \exp (-ik) \\
t_1 + t_2 \exp (ik) &
2t_{B}\cos k%
\end{array}%
\right) .
\end{eqnarray}
Alternatively, $h(k)$ can be expressed in the form
\[
h(k) =  h_I (k) I  + d (k) \cdot \sigma,
\]
with $I$ the identity matrix,  $\sigma=(\sigma_x,\sigma_y,\sigma_z)$ are Pauli matrices acting on the pseudospin $\psi_k$, $h_I = (t_A + t_B) \cos k  $, $d_x =  t_1 + t_2 \cos k $, $d_y=  t_2 \sin k  $ and $d_z=(t_A - t_B) \cos k $. Diagonalizing the Hamiltonian, we get the eigenvalues
\[
E(k) =(t_A + t_B) \cos k \pm \sqrt{\Delta}
\]
with
$\Delta=(t_A-t_B)^2\cos^2k + t_1^2 +  t_2^2 +2 t_1 t_2 \cos k$ .

\begin{figure}
\includegraphics[width=1.1\linewidth]{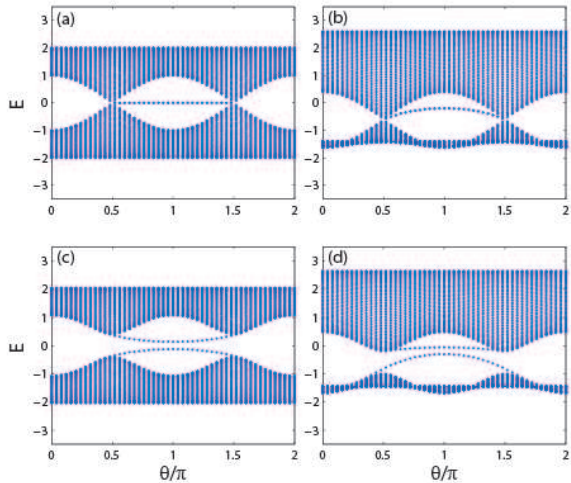}
\caption{(Color online) Energy spectrum for the extended SSH model with $100$ sites and
$\delta=0.5$ under OBCs. (a) $t_{A}=t_{B}=0$. (b) $t_{A}=t_{B}=0.3$. (c)
$t_{A}=-t_{B}=0.2$. (d) $t_{A}=0.5,t_{B}=0.1$.}
\label{Fig1}
\end{figure}

In the absence of NNN hopping terms, the spectrum of the SSH model is given by
$E = \pm \sqrt{ t_1^2 +  t_2^2 +2 t_1 t_2 \cos k}$,
which is composed of symmetric positive and negative branches. The two bands are separated by a band gap $\Delta E = 4 |\delta \cos \theta|$ at $k=\pi$ which is closed at $\theta= \pi/2$ and $3 \pi/2$ corresponding to $t_1=t_2$.
It is well known that the SSH model exhibits two topologically distinguishable phases protected by inversion
symmetry of the Hamiltonian. The two phases can be distinguished by the presence or absence of two-fold degenerate zero-mode edge states
under open boundary conditions (OBCs). As shown in Fig.1a for the system with $\delta=0.5$, the phase in the parameter regime $\theta \in (\pi/2 , 3 \pi/2)$ characterized by the presence of zero-mode edge states is topologically different from the phase in the regime $\theta \in (-\pi/2 , \pi/2)$. If we continuously
change the parameter $\theta$, a phase transition occurs at the $\theta=\pm \pi/2$.

The existence of zero modes in the SSH model is protected by both the inversion symmetry and particle-hole symmetry \cite{Ryu,symmetry}.
The presence of NNN hopping terms may break the symmetry. To see how the NNN hopping terms affect the properties of
the SSH model, we first consider the case with $t_A = t_B $. As shown in Fig.1b, the symmetric NNN hopping terms break the particle-hole symmetry,
but do not open the gap as the inverse symmetry is preserved. Next we consider the case with $t_A = - t_B $. While
such terms preserve the particle-hole symmetry, they break the inverse symmetry of the SSH model and thus open a gap as shown in Fig.1c. For a general case with $t_A \neq t_B$, the spectrum is split into two asymmetric bands with a gap opened.
Accompanying with opening of the gap, the edge modes split into two branches as displayed in Fig.1c and Fig.1d. As we continuously change the parameter $\theta$, the two branches of edge modes have no crossing and do not connect the separated bands, which indicates the system with the lower band filled to be a topologically trivial insulator.

\begin{figure}
\includegraphics[width=1.0\linewidth]{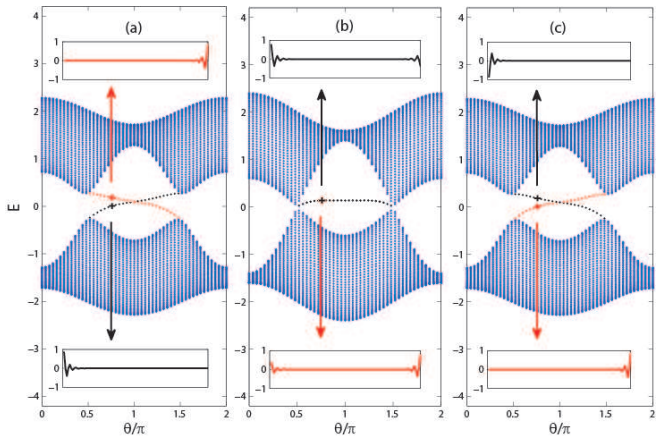}
\caption{(Color online) Energy spectrum for the extended SSH model
with $100$ sites, $\delta=0.5$ and $h=0.2$ under OBCs. (a), (b) and (c) correspond to
$\phi=-\pi/4$, $0$ and $\pi/4$, respectively. Insets show wavefuntions of edge modes.}
\label{fig2}
\end{figure}

As the fixed NNN hopping terms do not lead to topological nontrivial phenomena, we consider the model with both the NN and NNN hopping terms cyclically varying with the parameter $\theta$. To give a concrete example, we consider the modulation
\begin{eqnarray}
t_{A}=h \cos( \theta + \phi) , ~~~~ t_{B}= h \cos( \theta - \phi). \label{NNN1}
\end{eqnarray}
Here $h$ is a parameter to control the NNN hopping amplitude and an
additional parameter $\phi$ is introduced. Choosing a specific
$\phi$, we can get either symmetric or antisymmetric NNN hopping terms,
e.g., for $\phi=0$ we have $t_{A}=t_{B}$, and for $\phi=\pi/2$ we
get $t_{A}=-t_{B}$. In Fig.\ref{fig2}, we show the spectra of the
system versus $\theta$ under OBCs for several
typical parameters of $\phi$. As shown in Fig.\ref{fig2}a, the edge
modes exhibit nontrivial properties as they continuously connect the
bulk bands with the change of $\theta$. When we vary the
parameter $\phi$, we find that the edge mode spectrum exhibits
similar characters except for the case of $\phi=0$ as shown in
Fig.\ref{fig2}b, where the gap is closed and two branches of edge
modes merge together. Although systems with $\phi=\pi/4$ and
$\phi=-\pi/4$ have identical bulk spectra, we note that the edge
modes labeled by the same $E$ and $\theta$ locate at opposite edges
as displayed in Fig.\ref{fig2}a and Fig.\ref{fig2}c, which implies
the insulating states for systems with $\phi$ and $-\phi$ belonging
to different topological states.

\subsection{Topological invariant and phase diagram}

The gapless edge
states are generally associated with the nontrivial topological
properties of bulk systems. For the 1D extended SSH model with
cyclical modulation, we can define a Chern number
\begin{equation}
C=\frac{1}{2\pi}\int dkd\theta (\partial _{\theta}A_{k}-\partial
_{k}A_{\theta})
\end{equation}
in a 2D parameter space with $\theta$ being chosen
as the second parameter besides the momentum $k$,
where $A_k = i \langle \phi(k)|
\partial_k |\phi(k) \rangle $ and $A_\theta = i \langle \phi(k,\theta)|
\partial_\theta |\phi(k,\theta) \rangle $ with $\phi(k)$ the occupied Bloch
state. As a topological invariant, the Chern number can be used to characterize the topological properties of the family of the 1D system \cite{Thouless,TKNN}. It is well defined only when the gap of the bulk states
is fully opened throughout the Brillouin zone, and can be calculated numerically in
a discretized Brillouin zone \cite{Suzuki}. We find that the
Chern number of the half-filled extended SSH model with NNN hopping terms given by Eq.(\ref{NNN1}) is $-1$ when $\phi \in (-\pi,0)$,
whereas $C=1$ when $\phi \in (0,\pi)$. A phase transition occurs at the boundary of $\phi=0$ and $\pm \pi$. We note that taking different values of $h$
shall not change the phase diagram, however one needs to choose a suitable $h$ to guarantee the two bands never overlapping together and separated by a
finite gap.
\begin{figure}
\includegraphics[width=1.0\linewidth]{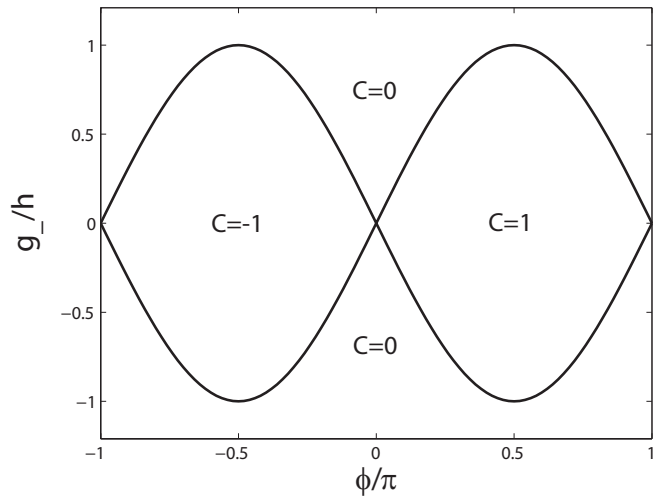}
\caption{Phase diagram for the extended SSH model with the NNN hopping amplitude given by Eq.(3). The number $C$ indicates the Chern number of
the lower band.} \label{fig3}
\end{figure}

Next we consider a more general case with NNN hopping amplitudes modulated as
\begin{eqnarray}
t_{A}=g_A + h \cos( \theta + \phi) , ~~~~ t_{B}= g_B + h \cos( \theta - \phi), \label{NNN2}
\end{eqnarray}
where $g_A$ and $g_B$ are two constants. For convenience, we define $g_{+} = (g_A + g_B)/2$ and $g_{-} = (g_A - g_B)/2$ and choose $h>0$. In order to guarantee the Fermi level lying in a gap between two bands, we need choose suitable parameters of $g_A$, $g_B$ and $h$ to avoid the overlap of the lower and upper bands.
Introducing two additional parameters produces a much richer phase diagram than the case with modulation described by Eq.(\ref{NNN1}).
In Fig.\ref{fig3}, we display the phase diagram in the parameter space of $\phi$ and $g_{-}/h$. There are altogether three topologically different phases, which can be characterized by the topological invariant $C=1$, $-1$ and $0$.
For a topological nontrivial system, the Chern number for each band
is an invariable integer unless the bulk gap closed \cite{Haldane},
therefore the phase boundary of different
topological phases can be determined by the gap closing point. For the extended SSH model with modulation of (\ref{NNN2}),
the eigen-energy is given by
$
E=2(g_{+} +  h \cos\theta\cos\phi) \cos k \pm\sqrt{\Delta}
$
with $ \Delta=4\cos^{2}k (g_{-}- h \sin\theta\sin\phi)^{2}+2(1+\cos k)+ 2(1-\cos k)\delta^{2}\cos^{2}\theta$.
The gap closes when $\Delta =0$, which enforces the condition
\begin{equation}
{g_{-}}/{h}= \pm \sin \phi \label{phaseboundary}
\end{equation}
for $k=\pi$ and $\theta=\pm {\pi}/{2}$.
Around the phase boundary, one can analytically derive the Chern number of the lower filled band
given by
\begin{equation}
C = \frac{1}{2} [ sgn(m_+) - sgn(m_-)]
\end{equation}
with $m_{\pm} = 2(g_{-} \pm h \sin \phi)$.
The numerical calculation of the Chern number also gives
exactly the same phase diagram as shown in Fig.3 with the phase boundary determined by Eq.(\ref{phaseboundary}). It is interesting that the phase diagram of our 1D extended SSH model has a similar structure as that of the 2D Haldane model \cite{Haldane}.

\begin{figure}
\includegraphics[width=1.0\linewidth]{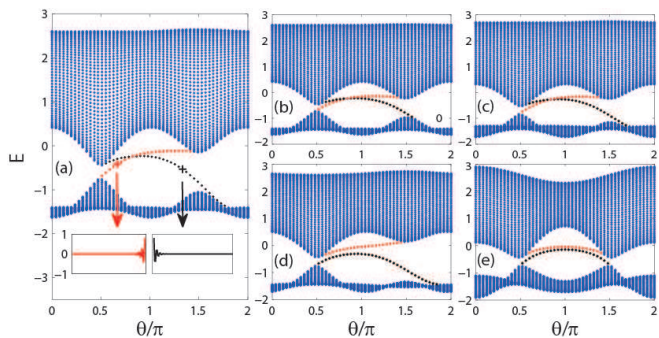}
\caption{(Color online) Energy spectrum of the extended SSH model with
$100$ sites and $\delta=0.5$ under OBCs. (a)
$g_{+}=0.3,h=0.15,g_{-}/h=0.5,\phi=0.5\pi$. Each of (b)-(e) has only
one parameter different from (a). (b) $h=0.1$. (c) $g_{+}=0.4$. (d)
$g_{-}/h=1.5$. (e) $\phi=0$. Insets show wavefuntions of edge modes.}
\label{fig4}
\end{figure}

The phase boundary is only associated with $\phi$ and the ratio of
$g_{-}$ and $h$, but is irrelevant to the value of $g_+$ and $h$.
As long as we keep $\phi$ and
$g_{-}/ h $ unchanged, varying $g_+$ and $h$ will not
affect the phase diagram. To give examples, we show spectra of systems with $g_-/h = 0.5$ and $\phi=0.5 \pi$ but different $h$ and $g_+$ under OBCs
in Fig.\ref{fig4}(a)-(c). Despite of the shape of spectra having some minor differences, the existence of continuous edge modes connecting the lower band and upper band for all these systems indicates that they are topologically nontrivial. As a contrast, we show the spectra of topologically trivial states in Fig.\ref{fig4}(d)-(e), where the upper and lower branches of edge modes are separated by a gap.

\subsection{Experimental proposal for observation of topological edge states}
The topological property of this model may be observed through the
experimental setup composed of series of single mode waveguides
arranged in a circular arc (Fig.5). As shown in Fig.5a, each dot
represents a waveguide vertical to the paper which corresponds to a
site in the tight binding Hamiltonian \cite{Kraus}. The propagating
light can tunnel from each waveguide to its neighboring waveguides,
and the hopping amplitude is inversely proportional to the distance
between them. The red solid dots in Fig.5 represent sites A, while
the green hollow ones are for sites B. Each of them are arranged in
a circular arc with an independent radius $r_{1}$ and $r_{2}$. The
angle between two neighbor red (or green) dots $\beta$ is set to a
constant, hence we can control the NNN hopping amplitudes $t_{A}$
and $t_{B}$ by adjusting $r_{1}$ and $r_{2}$, respectively. The distance
between two neighbor dots of different circular arc is determined by
$\alpha$, $r_{1}$ and $r_{2}$ all together, thus the NN hopping
amplitudes $t_{1}$ and $t_{2}$ can not be adjusted independently
from $t_{A}$ and $t_{B}$. However, the alternating relation between
$t_{1}$ and $t_{2}$ can be achieved by varying $\alpha$
individually. Hence we can control the hopping amplitudes $t_1, t_2,
t_{A}, t_{B}$ by varying $\alpha, \beta, r_{1}, r_{2}$ in Fig.5a to
design the waveguide system, which can be described by the
Hamiltonian (1). By choosing proper parameters, we can prepare the
system in either topologically nontrivial or trivial state as shown
in Fig. {\ref{fig4}}(a) or (d). If we take parameters of
Fig.{\ref{fig4}}(a) with $\theta=-2\pi/3$ (equivalently
$\theta=4\pi/3$ as marked in Fig.{\ref{fig4}}(a)), one can observe
the edge states located in the left edge by measuring intensity
distributions of injecting light as a function of the position of
the waveguides \cite{Kraus}. If we change $\theta$ adiabatically in
the direction of light propagation, in a topological nontrivial
case, one can observe the edge state localized in one end of the
lattices pumping to the other side as we vary the $\theta$ from
$-2\pi/3$ to $2\pi/3$, while in the topological trivial case (e.g.
Fig.{\ref{fig4}}(d)) the edge state will always localize in the same
side.
\begin{figure}
\includegraphics[width=0.9\linewidth]{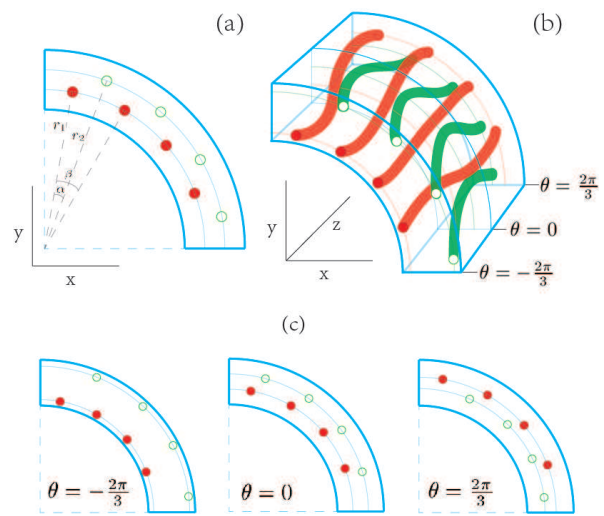}
\caption{(Color online) Red circles stand for site A and green ones stand for site
B in the model. (a) Schematic diagram for the arrangement of waveguides for a given $\theta$. (b) Along the propagation direction, $\theta$ can be varied along the length of the waveguide by smoothly changing the spacing between successive waveguides. (c) Examples for the geometrical arrangement of waveguides for several $\theta$.} \label{fig5}
\end{figure}

To study the feasibility of this experimental proposal, we add a
random disorder perturbation $t_d $ to each hopping amplitude to check the robustness
of the topological edge states under the disorder perturbation, where $t_d$ randomly distributes in the range of $[-W/2,W/2]$.
Taking the system with parameters as
given in Fig.{\ref{fig4}}(a) with $\theta=4\pi/3$, we find the edge state located in the left boundary is robust
even the disorder amplitude takes $W=0.2$.

\subsection{Other models}

Finally, we consider an alternative model which also exhibits the same phase diagram as model (\ref{H1}) with NNN hopping amplitudes given by Eq.(\ref{NNN2}).  The model can be described by
\begin{equation}
H=H_{SSH}+H_{\mu}
\end{equation}
with $H_{SSH}$ the Hamiltonian of SSH model and $H_{\mu}$ given by
\begin{equation}
H_{\mu}=\sum\limits_{n}^{N}\mu_{A}c_{n,A}^{+}c_{n,A}+\mu_{B}c_{n,B}^{+}c_{n,B},
\end{equation}
where the alternating on-site potentials $\mu_A$ and $\mu_B$ are given by
\begin{eqnarray}
\mu_{A}=g_A + h \cos( \theta - \phi) , ~~~~ \mu_{B}= g_B + h \cos(
\theta + \phi). \label{mu}
\end{eqnarray}
The Hamiltonian in momentum space can be represented as $H = \psi^{\dagger}_k h(k) \psi_k$ with
\begin{eqnarray}
h(k)=
\left(
\begin{array}{cc}
 \mu_{A} & t_1 + t_2 \exp (-ik) \\
t_1 + t_2 \exp (ik) &
 \mu_{B}%
\end{array}%
\right),
\end{eqnarray}
The spectrum of the system is given by
\[
E(k)=g_{+}+h\cos\theta\cos\phi\pm\sqrt{\Delta}
\]
with $ \Delta=(g_{-}+ h\sin\theta\sin\phi)^{2}+2(1+\cos k) +2(1-
\cos k)\delta^{2}\cos^{2}\theta. $ It is easy to find the phase
boundary given by $ {g_{-}}/{h}= \pm \sin \phi$ corresponding to the
close of energy gap for $k=\pi$ and $\theta=\mp {\pi}/{2}$.
Following similar procedures for model (\ref{H1}), we can find
that the phase diagram for the above SSH model with alternating
on-site potentials is also given by Fig.\ref{fig3}. The above model
is possible to be realized in cold atomic systems trapped in optical
superlattices \cite{Foelling,Atala}.

In order to see the connection of the 1D model to the 2D Haldane model, we consider the model with both NNN hopping terms and
modulated on-site potentials described by the following Hamiltonian
\begin{equation}
H=H_{SSH}+H_{NNN}+H_{\mu}. \label{H-both}
\end{equation}
Similarly, the Hamiltonian in momentum space can be represented as $H = \psi^{\dagger}_k h(k) \psi_k$ with
\begin{eqnarray}
h(k)=
\left(
\begin{array}{cc}
2t_{A}\cos k +\mu_{A} & t_1 + t_2 \exp (-ik) \\
t_1 + t_2 \exp (ik) &
2t_{B}\cos k+ \mu_{B}%
\end{array}%
\right). \label{both}
\end{eqnarray}
Diagonalizing the Hamiltonian, we get the eigenvalues
\[
E(k) =(t_A + t_B) \cos k + (\mu_{A}+\mu_{B})/2 \pm \sqrt{\Delta}
\]
with $\Delta=[\cos k(t_A-t_B)+(\mu_A-\mu_B)/2]^2 + t_1^2 +  t_2^2 +2
t_1 t_2 \cos k$. If both the NNN hopping amplitudes ($t_A$, $t_B$) and alternating on-site potentials ($\mu_A$, $\mu_B$) are modulated in a similar way as Eq.(\ref{NNN2}) or Eq.(\ref{mu}), topological nontrivial phases characterized by nonzero Chern numbers can also be found.

For the convenience of later comparison, here we briefly review the Haldane model. As discussed in details in Ref.\cite{Haldane,Vanderbilt,Hao}, the lattice
tight-binding Hamiltonian of the Haldane model is given by
\begin{eqnarray}
H=\sum_{i} \tau_{0} c_{i}^{\dagger}c_{i}+\sum_{\langle
i,j\rangle} \tau_{1} c_{i}^{\dagger}c_{j}+\sum_{\langle\langle
i,j\rangle\rangle} \tau_{2} e^{i\phi_{i,j}}c_{i}^{\dagger}c_{j} ,
\end{eqnarray}
where the summation $\sum_{i}$ is defined in the 2D honeycomb lattice, which is composed of two sublattices labeled by ``A" and ``B", respectively. Here
$\tau_{0}=M$ for site A and $\tau_{0}=-M$ for site B, $\tau_{1}$ denotes the nearest-neighbor hopping amplitude and $\tau_{2}$ the NNN hopping amplitude. The magnitude of the
phase is set to be $|\phi_{i,j}|=\phi$, and the direction of the
positive phase is clockwise following Haldane's work\cite{Haldane}.
Choosing the zigzag edges in the $x$ direction \cite{Hao,note} and  using the basis of two component ``spinors" $\psi_{K}^\dagger = (c_{A,K}^{\dagger}, c_{B,K}^{\dagger})$ ($K=(k_x,k_y)$) of Bloch states constructed on the two sublattices, we can write the Hamiltonian in the momentum space as:
\begin{eqnarray}
h(K)= \left(
\begin{array}{cc}
2s_{+}\cos k_{x} +u_{+} & r + \tau_1 \exp (-ik_{x}) \\
r + \tau_1 \exp (ik_{x}) &
2s_{-}\cos k_{x} +u_{-}%
\end{array}%
\right), \label{h-haldane}
\end{eqnarray}
where $s_{\pm}=2\tau_{2}\cos (\frac{\sqrt{3}k_{y}}{2} \pm\phi )$,
$u_{\pm}=\pm M + 2 \tau_{2}\cos (\sqrt{3}k_{y} \mp\phi )$ and
$r = 2 \tau_{1}\cos \frac{\sqrt{3}k_{y}}{2}$. The form of this
Hamiltonian is very similar to Eq.(\ref{both}), and with similar
procedure we can calculate the gap closing conditions, which give the
phase transition boundaries of the Haldane model determined by $M/\tau_2 = \pm  3\sqrt{3} \sin \phi$. The Chern number of
the lower filled band can also be analytically given by $C =
\frac{1}{2} [ sgn(m_+) - sgn(m_-)]$, where the effective mass
becomes $m_{\pm}=M\pm 3\sqrt{3} \tau_{2} \sin \phi$.

Comparing Eq.(\ref{both}) to Eq.(\ref{h-haldane}), if we take $t_1 = 2 \tau_{1}\cos \frac{\sqrt{3} \theta}{2}$, $t_2 = \tau_1$, $t_{A,B}=2\tau_{2}\cos (\frac{\sqrt{3} \theta}{2} \pm\phi )$ and $\mu_{A,B}=\pm M + 2 \tau_{2}\cos (\sqrt{3}\theta \mp \phi )$, we observe that the model of (\ref{H-both}) can be mapped to the Haldane model by identifying $k$ to $k_x$ and $\theta$ to $k_y$. The above mapping gives us a straightforward understanding of the connection of the 1D generalized SSH model with fine tuning parameters to the well-known 2D Haldane model. Similarly, a mapping between the 1D superlattice and quasi-periodic systems and the 2D Hofstadter problems has been well established \cite{Shu,Kraus,Madsen}. However, we note that the model described by Eq.(\ref{both}) with artificially fine tuning parameters is hard to be experimentally realized as both the hopping parameters and alternative on-site potentials need to be tuned simultaneously. On the other hand, models with only either the modulated NNN hopping terms or modulated on-site potentials also exhibit the nontrivial phase diagram similar to that of Haldane's model, but are much easier to be realized in experiments.

\section{ Summary}
In summary, we study the extended SSH model with
both the NN and NNN hopping terms periodically modulated by a
parameter $\theta$. Our results show that the family of extended SSH
model supports topologically nontrivial edge modes with respect to
the modulation parameter $\theta$. The topologically nontrivial
phase diagram shows some surprising connections between the 1D
extended SSH models and the 2D Haldane model. We propose an experimental setup
composed of photonic crystals to realize our model system. Some
other models with similar phase diagrams are also discussed and
their connection to the 2D Haldane model is addressed.

\begin{acknowledgments}
This work has been supported by National Program for Basic Research
of MOST, NSF of China under Grants No.11374354, No.11174360
and No.10974234, and 973 grant.
\end{acknowledgments}


\begin{thebibliography}{99}
\bibitem{symmetry} Here the inverse symmetry implied $h(k) = \sigma_x h(-k) \sigma_x$, and the particle-hole symmetry means
that the spectra are symmetry about $E=0$, i.e., if $E$ is an eigenvalue of $H$, then so is $-E$.
\bibitem{note} The direction of the Zigzag edges in this work is different from that of Ref.{\cite{Hao}},
where the direction is along the y direction as shown in the Fig.1(a) therein. Replacing $k_x$ to $k_y$ and $k_y$ to $k_x$ in our work, we can get results
in Ref.{\cite{Hao}}.
\end{thebibliography}
\end{document}